\documentclass[preprint,12pt,authoryear]{elsarticle}
\usepackage{amsthm}
\usepackage{array}
\usepackage{graphicx}
\usepackage{float}
\usepackage{booktabs}
\usepackage{subcaption}
\usepackage{tikz}
\usetikzlibrary{positioning,decorations.pathreplacing}

\usepackage{amssymb}
\usepackage{amsmath}

\journal{Nuclear Physics B}
\theoremstyle{definition}
\newtheorem{defn}{Definition}

\begin{document}

\begin{frontmatter}



\title{On feature selection in double-imbalanced data settings: a Random Forest approach} 


\author{Fabio Demaria\corref{cor1}} 

\cortext[cor1]{Email: fabio.demaria@unimore.it}
\affiliation{organization={Department of Economics ``Marco Biagi''},
            addressline={Via J. Berengario, 51}, 
            city={Modena},
            postcode={41121}, 
            state={Italy}}

\begin{abstract}
Feature selection is a critical step in high-dimensional classification tasks, particularly under challenging conditions of \textit{double imbalance}, namely settings characterized by both class imbalance in the response variable and dimensional asymmetry in the data (\(n \gg p\)). In such scenarios, traditional feature selection methods applied to Random Forests (RF) often yield unstable or misleading importance rankings. This paper proposes a novel thresholding scheme for feature selection based on minimal depth, which exploits the tree topology to assess variable relevance. Extensive experiments on simulated and real-world datasets demonstrate that the proposed approach produces more parsimonious and accurate subsets of variables compared to conventional minimal depth-based selection. The method provides a practical and interpretable solution for variable selection in RF under double imbalance conditions.
\end{abstract}




\end{frontmatter}

\section{Introduction}
\label{sec:1}
Class imbalance is a prevalent issue in machine learning, occurring when one class is significantly underrepresented relative to others in the target variable. This asymmetry, commonly quantified by the imbalance ratio (IR), poses substantial challenges for standard classification algorithms, which tend to be biased toward the majority class. Class imbalance is especially critical in high-stakes applications such as fraud detection, medical diagnostics, and risk assessment, where identifying the minority class accurately is crucial \citep{He2009, Lopez2013}.
This study introduces and addresses a more complex condition we refer to as \textit{double imbalance} -- a scenario in which the dataset is not only affected by class imbalance but also exhibits a dimensional asymmetry between the number of units and variables (see Section \ref{sec:2} for a formal definition), typically characterized by high-sample/low-dimension configurations (\(n \gg p\)). This condition can distort tree-based learning, especially in Random Forests (RF), by promoting overly deep trees and biased feature importance metrics such as minimal depth \citep{Strobl2007, Ishwaran2010}.
To address class imbalance, the literature has proposed a range of solutions, including data-level methods (e.g., SMOTE, under-sampling), algorithm-level adaptations (e.g., cost-sensitive learning, adjusted decision thresholds), and hybrid strategies \citep{Galar2012, Krawczyk2016, He2009}. Among algorithm-level approaches, RF-based models, such as Balanced Random Forests (BRF) and the Random Forest Quantile (RFQ) classifier, have shown promising results, especially when the classification threshold is tuned to favor the minority class \citep{OBrien2019}.
Building on this line of research, we propose a novel algorithm-level contribution specifically tailored to double imbalance scenarios. Our method enhances variable selection by modifying the threshold used in minimal depth distributions to reflect both class imbalance and dimensional asymmetry. We introduce a data-driven adjustment factor that regularizes the minimal depth threshold as a function of the \(n/p\) ratio, mitigating overfitting and improving the detection of informative features in double imbalanced data settings.

The key contribution of this study is methodological: we develop and validate a new thresholding scheme for feature selection in RF that is robust to double imbalance settings. Through empirical evaluations on real-world datasets and synthetic data, we demonstrate that the proposed approach yields more parsimonious and accurate variable subsets compared to conventional minimal depth-based selection.

This work is structured as follows. Section~\ref{sec:2} formally defines the class imbalance problem and reviews the main approaches developed to address it. The concept of double imbalance is also introduced in this section. In Section~\ref{sec:3}, we present a novel thresholding scheme for identifying the most important variables in RF using the minimal depth concept. Section~\ref{sec:4} evaluates the effectiveness of the proposed method through a comparative analysis on real-world datasets characterized by double imbalance and validates its performance using simulated data. Finally, Section~\ref{sec:5} summarizes the key findings and concludes the study.

\section{Class imbalance data setting}
\label{sec:2}
Class imbalance, a well-known challenge in machine learning, arises when the number of observations is unevenly distributed across classes, with the minority class being significantly underrepresented \citep{He2009, Lopez2013}. To formally define the imbalanced data setting, let the learning data be denoted by \(\mathcal{L} = \{(X_i, Y_i)\}\) with \(i=1,\ldots,n\), where \(X_i \in \mathbf{X}\) represents the \(p\)-dimensional feature vector and \(Y_i \in \{0, 1\}\) is the binary ordinal response. It is assumed that \((X_i, Y_i)\) are independently and identically distributed (i.i.d.) from a common distribution \(\mathbb{P}\). The degree of imbalance is quantified using the Imbalance Ratio (IR), defined as:
\begin{equation}
\text{IR} = \frac{C_0}{C_1},
\end{equation}
where \(C_0\) and \(C_1\) denote the number of majority and minority class samples, respectively. A dataset is considered imbalanced if \(\text{IR} \gg 1\).\\

\begin{defn}
Class imbalance is characterized by an imbalance ratio: \(\text{IR} = \frac{C_0}{C_1}\), where \(C_0\) and \(C_1\) are the number of observations in the majority and minority classes, respectively. A dataset is said to be \textit{imbalanced} if \(\text{IR} \gg 1\).\\
\end{defn}

Many real-world classification problems are affected by a pronounced class imbalance, where one or more classes are significantly underrepresented compared to others. This issue is particularly prevalent in domains such as fraud detection, medical diagnostics, risk assessment, and information filtering \citep{Lopez2013, Blaszczynski2015}. Class imbalance presents a challenge because standard learning algorithms are typically biased toward the majority class, often resulting in poor recognition performance on the minority class \citep{Sun2009, Lopez2013}.

Despite the proliferation of methods to mitigate this challenge, identifying the appropriate strategy for specific data conditions remains an open area of research \citep{Napierala2016}. The performance degradation in such settings is not solely due to the imbalance ratio, but is often exacerbated by overlapping class distributions, small disjuncts, or the presence of noisy data \citep{Jo2004, Prati2004, Napierala2010}. Thus, addressing class imbalance is a multifaceted problem requiring a nuanced understanding of the data characteristics.

\subsection{Types of class imbalance}
Class imbalance can be categorized into two types: \textit{marginal} and \textit{conditional} imbalance \citep{Fithian2014}. Marginal imbalance refers to the scenario where the overall prior probability of one class is significantly lower than that of another, such as in datasets with \( P(Y=1) \approx 0 \). This is common in rare event prediction tasks. Conditional imbalance, on the other hand, occurs when the class conditional probabilities are skewed for most input values, e.g., \( P(Y=1\mid X=0) \approx 0 \) and \( P(Y=1\mid X=1) \approx 1 \).

Moreover, a distinction is made between \textit{intrinsic} and \textit{extrinsic} imbalance \citep{He2009}. Intrinsic imbalance arises from the nature of the problem space itself, while extrinsic imbalance results from external constraints such as temporal limitations or storage issues. Both types may adversely affect model learning.

\subsection{Safe vs. unsafe examples}
The classification difficulty of examples in imbalanced datasets can vary greatly depending on their location in the feature space. Safe examples are typically situated in homogeneous regions where most neighboring instances share the same label, making them easier for classifiers to learn. In contrast, unsafe examples lie in more ambiguous regions and are more susceptible to misclassification \citep{Laurikkala2001, Blaszczynski2015}.
Unsafe examples can be further divided into three categories \citep{Napierala2016}. \textbf{Borderline examples} are positioned near the decision boundary between classes. \textbf{Rare examples} refer to isolated small clusters of minority instances embedded in the majority class region, yet not sufficiently remote to be considered outliers. Finally, \textbf{Outliers} are distant minority instances that may reflect rare but valid sub-concepts not well represented in the training data \citep{Kubat1997}.
Assessing the nature of these examples often involves analyzing the local neighborhood using $k$-nearest neighbors to evaluate class distribution. Visualization techniques such as multidimensional scaling (MDS) or t-SNE are also employed to distinguish example types in low-dimensional projections \citep{VanDerMaaten2008, Napierala2012}.

\subsection{Methods to deal with class imbalance}
In this section we focus on the most related studies concerning the properties of imbalanced data and their consequences for learning classifiers or pre-processing methods. For a more comprehensive review of various methods proposed to deal with class imbalance, the reader is referred to He and Garcia (2009), He and Ma (2013).
There are three primary methodological categories designed to mitigate the adverse effects of class imbalance \citep{He2009, Galar2012, Lopez2013, Krawczyk2016, Das2018}:

\paragraph{\bf Data-level methods:} These techniques involve modifying the dataset to achieve a more balanced class distribution. Over-sampling increases the number of minority instances, often through duplication or synthetic generation (e.g., SMOTE), while under-sampling removes instances from the majority class. These strategies aim to make standard algorithms more effective by reducing class imbalance before training.

\paragraph{\bf Algorithm-level methods:} This approach involves modifying the learning algorithm itself to reduce its bias toward the majority class. Examples include altering the decision thresholds, changing the cost functions, or incorporating ensemble learning strategies specifically tuned for imbalanced data.

\paragraph{\bf Cost-sensitive learning:} These methods combine data and algorithm-level strategies by incorporating varying misclassification costs for each class into the learning process. This ensures that misclassifying minority class instances is penalized more heavily, thus encouraging the classifier to give them more attention during training \citep{LoyolaGonzalez2016}.\\

Overall, traditional classifiers such as decision trees or random forests are typically designed for balanced class distributions, and their performance can degrade significantly in imbalanced settings \citep{OBrien2019}. To evaluate classifiers under imbalance, traditional accuracy metrics are insufficient; alternative metrics such as precision, recall, F1-score, ROC curves, and precision-recall curves are recommended \citep{He2009}.

\subsection{Double imbalance: class and dimensional asymmetry}
In addition to the well-known issue of class imbalance, many practical datasets also exhibit a second structural asymmetry -- what we refer to as \textit{double imbalance}. This condition arises when both (i) the distribution of the dependent variable is highly skewed (class imbalance), and (ii) there is a disproportionate ratio between the number of observations and the number of features, typically in high-sample/low-dimension settings where \( n \gg p \). While larger sample sizes are generally favorable for learning, in the context of Random Forests and other tree-based ensembles, this can lead to the construction of excessively deep trees. Deep trees increase the risk of overfitting and can distort feature importance measures such as minimal depth, which may artificially favor variables that appear early in the tree splits \citep{Strobl2007, Ishwaran2010}. Furthermore, this structural imbalance may exacerbate biases inherent to class-imbalanced learning by reinforcing dominant patterns present in the majority class. Addressing double imbalance thus requires methodological adaptations that consider both the target class distribution and the geometry of the feature space.\\

\begin{defn}
A dataset is said to exhibit \textbf{double imbalance} if, in addition to a high imbalance ratio (\(\text{IR} \gg 1\)), it also satisfies a dimensional asymmetry in its design matrix such that the number of observations is much greater than the number of features, i.e., \(n \gg p\). \\
\end{defn}

In this case, the dataset not only suffers from skewed class distribution, but also from a high-sample/low-dimension configuration that may bias variable selection and promote overfitting in deep decision tree models.

\section{Algorithm-level methods: Random Forest-based models}
\label{sec:3}
Random Forests (RF) are a powerful ensemble learning method based on decision trees, using bagging and random feature selection \citep{Breiman2001}. However, like many classifiers, they can be biased toward the majority class when faced with imbalanced data \citep{Galar2012}.

\subsection{Balanced Random Forest}
Balanced Random Forests (BRF) modify the bootstrap sampling procedure by drawing a balanced subset of the training data for each tree \citep{Chen2004}. Specifically, each tree is built from a bootstrap sample containing all \(N_{min}\) instances from the minority class and an equal number of randomly selected instances from the majority class \citep{Liaw2002}:
\begin{equation}
\text{BRF Sample} = \text{Minority} (N_{min}) + \text{Random Majority} (N_{min})
\end{equation}
This balanced sample is then used to grow a decision tree using standard RF procedures. The method aims to correct bias without modifying the RF algorithm itself.

\subsection{Random Forest Quantile classifier}
\label{sec:rfq}
The Random Forest Quantile (RFQ) classifier addresses class imbalance by adapting the decision threshold using estimated conditional probabilities \citep{OBrien2019}. The RFQ classifier assigns a new observation \(x\) to the minority class if:
\begin{equation}
\delta_{RFQ}(x) = 1\{\hat{p}_{RF}(x) \geq \pi\},
\end{equation}
where \(\hat{p}_{RF}(x)\) is the class probability estimated by the RF ensemble, and \(\pi\) is the prevalence of the minority class in the training data \citep{OBrien2019}. This shifts the decision boundary to enhance sensitivity to the minority class.

The RFQ classifier achieves optimality in terms of maximizing the sum of true positive rate (TPR) and true negative rate (TNR):
\begin{equation}
\text{TPR} = \frac{TP}{TP + FN}, \quad \text{TNR} = \frac{TN}{TN + FP}.
\end{equation}
This approach improves classification fairness and performance under imbalance by accounting for the skewed prior.

\subsection{Variable selection in RF}
\label{sec:importance}
Traditional approaches for estimating feature importance in RF include the mean decrease in impurity (MDI) and permutation importance, also known as Variable Importance (VIMP) or Breiman-Cutler importance \citep{Breiman2001, Hapfelmeier2014}. VIMP quantifies a variable's importance by evaluating the increase in out-of-bag (OOB) prediction error after permuting that variable, thereby disrupting its association with the outcome \citep{Breiman2001}.
While widely used, these importance metrics have limitations in class-imbalanced settings—especially when applied to models like the RFQ classifier, which are designed to optimize decision boundaries using probability estimates rather than error minimization. Since VIMP relies directly on prediction error, it may fail to accurately reflect the relevance of variables when class distributions are imbalanced. As highlighted by \citet{OBrien2019}, error-based importance metrics can be misleading under imbalance, potentially overestimating the contribution of variables that favor the majority class.

An alternative, more robust approach for feature selection in this context is the \textit{minimal depth} method. Rather than depending on predictive performance, minimal depth evaluates a variable's position within the tree structure. The underlying rationale is that variables selected closer to the root node are more influential in partitioning the data and thus hold greater predictive power \citep{Ishwaran2010, Ishwaran2011}. This method is advantageous because it is model-agnostic within the class of tree ensembles, does not require OOB predictions, and is applicable to both supervised and unsupervised forests.
Formally, minimal depth is defined through the concept of a variable’s maximal subtree. For a given variable \(v\), the first-order maximal subtree is the largest subtree for which \(v\) appears as the splitting variable at the root. Let \(D(T)\) be the maximum depth of tree \(T\), and let the depth \(d\) of a node represent its distance from the tree root, where \(d \in \{0, 1, \ldots, D(T)\}\). The minimal depth \(D_v\) is then defined as the depth of the root node of the nearest maximal subtree in which \(v\) initiates a split. A smaller \(D_v\) indicates higher importance, as it suggests that \(v\) plays a decisive role early in the decision-making process.
In summary, minimal depth offers a principled, structure-based alternative to error-driven importance measures, and is particularly well-suited to the challenges posed by imbalanced classification tasks.

\subsection{A new thresholding scheme for variable selection in RF}
\label{sec:new_threshold}
In this study, we propose a modification to the minimal depth distribution to address cases with a large number of units \( n \) relative to the number of variables \( p \).  This new threshold represents a desirable regularization effect to avoid overfitting in high-sample/low-dimension scenarios. Specifically, to adjust for the imbalance between the sample size and the number of variables, we introduce an \textit{adjustment factor}, defined as:
\begin{equation}
\label{eq:adj_factor} \\
\psi = \frac{1}{\lambda} \times \ln \left( \frac{n}{p} \right)
\end{equation}
where \( \lambda = \sqrt{\frac{n}{p}} + \ln(p) \) represents a scaling parameter. Specifically, the term $\sqrt{n/p}$ captures the relative sample size per parameter and increases with $n \gg p$, thereby accounting for sample efficiency. The term $\log(p)$ introduces a soft penalty for model complexity, increasing slowly as the number of predictors $p$ grows.
Thus, \(\psi\) acts as a non-linear shrinkage term: decreases as the sample size $n$ increases relative to $p$, and moderates over-penalization when $p \approx n$. This mechanism therefore adapts automatically to the data regime. In other words, this factor scales the probability distribution to reflect that in large datasets with many observations (\( n \)), the likelihood of a variable being selected for early splits should decrease. 

For each depth \( d \), we compute the distribution based on the number of nodes at that depth. Let \( L_d \) be the cumulative number of nodes from depth 0 to depth \( d-1 \) and \( \ell_d \) the number of nodes at depth \( d \).
The probability that the variable does not split at depth \( d \) is logarithmically modeled as: 
\begin{align}
P_1 = L_d \times \log\left(1 - \frac{1}{p^{\ast}}\right) \\
P_2 = \ell_d \times \log\left(1 - \frac{1}{p^{\ast}}\right) 
\end{align}
where \( p^{\ast} = p \times \psi \) is the scaled number of variables. This scaling modifies the original minimal depth probability distribution, accounting for the influence of large sample sizes on the likelihood of variable selection at various depths, allowing for setting dynamic thresholds during variable selection.
Consequently, the probability that a variable \( v \) first splits at depth \( d \) is given by: 
\begin{equation}
\mathbb{P}(D_v = d) = \exp(P_1) \times \left[1 - \exp(P_2)\right] 
\end{equation}
This probability quantifies the likelihood that variable \( v \) initiates a split at depth \( d \), and this calculation is repeated for all depths \( d = 0, 1, \dots, D(T)-1 \).

Thus, building on the work of \citet{Ishwaran2010}, who proposed using the mean of the minimal depth distribution as a threshold for identifying relevant features, we adapt the minimal depth distribution for a weak variable \( v \) in settings where \( n \gg p \), resulting in the following expression: 
\begin{equation}
\label{eq:min_depth_distr}
\mathbb{P}(D_v = d \mid \ell_0, \ldots, \ell_{D(T)-1}, v \text{ is noisy}) = \left[ 1 - \left(1 - \frac{1}{p^{\ast}}\right)^{\ell_d} \right] \prod_{j=0}^{d-1} \left(1 - \frac{1}{p^{\ast}}\right)^{\ell_j} 
\end{equation}
where  \( \ell_d \) represents the number of non-terminal nodes at depth \( d \),  \( \prod_{j=0}^{d-1} \left(1 - \frac{1}{p^{\ast}} \right)^{\ell_j} \) accounts for the probability that \( v \) was not selected to split at any previous depth before \( d \), \( p^{\ast} \) is the adjusted variable count, reflecting the imbalance between the number of units \( n \) and variables \( p \).
This adjustment reflects the impact of a large sample size on the likelihood of selecting a variable earlier in the tree, thus favoring variables that split closer to the root in large datasets.

\section{Empirical results}
\label{sec:4}
\subsection{Comparison of RF models}
To demonstrate the effectiveness of the proposed method in identifying the most important variables in double imbalanced data settings, we selected three real-world datasets from the UCI repository, representing different domains, dataset sizes, and imbalance ratios (Table \ref{tab:data}). These datasets were chosen because they are commonly used in experimental studies on class imbalance \citep{Khoshgoftaar2011, Lopez2013, Blaszczynski2015}.

\begin{table}[ht]
\centering
\footnotesize
\caption{Data attributes.}
\renewcommand{\arraystretch}{1.2}
\begin{tabular}{l>{\centering\arraybackslash}m{2cm}>{\centering\arraybackslash}m{2cm}>{\centering\arraybackslash}m{2cm}>{\centering\arraybackslash}m{1.5cm}>{\centering\arraybackslash}m{1.5cm}}
\toprule
\textbf{Dataset} & \textbf{N. of examples} & \textbf{N. of features} & \textbf{Minority class} & \textbf{IR} & \textbf{DA$^\ast$} \\
\midrule
Adult & 48,842 & 15 & 0 & 3.19 & 3256.133 \\
Bank & 45,211 & 17 & no & 7.54 & 2659.471 \\
Coupon & 12,684 & 26 & 1 & 1.31 & 487.846 \\
\bottomrule
\end{tabular}
\caption*{\textbf{Note}: $^\ast$Dimensional asymmetry $(n/p)$.}
\label{tab:data}
\end{table}

We employed the RFQ and BRF classifiers described in Section~\ref{sec:3}, as both have been shown to be highly effective for imbalanced data settings \citep{OBrien2019, Chen2004}. Additionally, we compared these models with the standard RF classifier, which does not account for class imbalance. All models were implemented using forests with 5,000 trees grown on each training dataset, with trees grown to purity (\texttt{nodesize} = 1) and \texttt{mtry} = \( p/3 \). Results were averaged over 100 iterations and are reported in Table~\ref{tab:threshold_performance}.

\begin{table}[ht]
\centering
\footnotesize
\caption{Comparison of standard and adjusted minimal depth thresholds.}
\renewcommand{\arraystretch}{1.2}
\begin{tabular}{l>{\centering\arraybackslash}m{2cm}>{\centering\arraybackslash}m{2cm}>{\centering\arraybackslash}m{2cm}>{\centering\arraybackslash}m{2cm}}
\toprule
\textbf{Dataset} & \textbf{Std. threshold\(^1\)} & \textbf{N. selected features} & \textbf{Adj. threshold\(^2\)} & \textbf{N. selected features} \\
\midrule
\textbf{RFQ classifier} & & & & \\
Adult & 22.201 & 14 & 4.043 & 6 \\
Bank & 11.426 & 16 & 4.047 & 8 \\
Coupon & 7.751 & 23 & 4.857 & 15 \\
\midrule
\textbf{BRF} & & & & \\
Adult & 19.169 & 14 & 3.606 & 3 \\
Bank & 11.451 & 16 & 4.366 & 8 \\
Coupon & 7.256 & 23 & 4.077 & 10 \\
\midrule
\textbf{Standard RF} & & & & \\
Adult & 22.114 & 14 & 4.465 & 6 \\
Bank & 11.458 & 16 & 4.075 & 8 \\
Coupon & 7.746 & 23 & 4.821 & 15 \\
\bottomrule
\end{tabular}
\caption*{\scriptsize \textbf{Note(s)}: \(^1\)Minimal depth thresholding as defined in \citet{Ishwaran2008}. \(^2\)Adjusted threshold defined in Equation~\ref{eq:min_depth_distr}.}
\label{tab:threshold_performance}
\end{table}
The results indicate that the standard minimal depth threshold -- defined as the mean of the minimal depth distribution -- tends to retain nearly all variables across the different RF models in scenarios characterized by double imbalance. This behavior is consistent across all three datasets, where the number of selected features is close to the total number of available features. In contrast, the adjusted threshold, which incorporates a data-driven scaling factor into the minimal depth distribution (see Equation~\ref{eq:min_depth_distr}), yields substantially more parsimonious subsets of variables. This result suggests that the proposed method is more effective in isolating the most informative features under complex distributional conditions.

A graphical comparison of the RFQ and BRF models using the dataset-specific adjusted minimal depth threshold is shown in Figures~\ref{fig:adult}, \ref{fig:bank}, and \ref{fig:coupon}. As can be seen from the pairwise comparisons, the proposed threshold performs effectively in both models.

\begin{figure}[H]
\caption{Minimal depth thresholding for dataset `Adult' (UCI).}
  \begin{subfigure}{0.46\textwidth}
      \scalebox{0.45}{%

}
    \caption{RFQ classifier.} \label{fig:adult_rfq}
  \end{subfigure}%
  \hspace*{\fill}
  \begin{subfigure}{0.46\textwidth}
      \scalebox{0.45}{%
%
}
    \caption{BRF model.} \label{fig:adult_brf}
  \end{subfigure}
    \caption*{\footnotesize \textbf{Note}: Adjusted minimal depth threshold (Eq. \ref{eq:min_depth_distr}). Averaged results over 100 runs.}
    \label{fig:adult}
\end{figure}

\begin{figure}[H]
\caption{Minimal depth thresholding for dataset `Bank' (UCI).}
  \begin{subfigure}{0.46\textwidth}
      \scalebox{0.45}{%
%
}
    \caption{RFQ classifier.} \label{fig:bank_rfq}
  \end{subfigure}%
  \hspace*{\fill}
  \begin{subfigure}{0.46\textwidth}
      \scalebox{0.45}{%
   %
}
    \caption{BRF model.} \label{fig:bank_brf}
  \end{subfigure}
    \caption*{\footnotesize \textbf{Note}: Adjusted minimal depth threshold (Eq. \ref{eq:min_depth_distr}). Averaged results over 100 runs.}
    \label{fig:bank}
\end{figure}

\begin{figure}[H]
\caption{Minimal depth thresholding for dataset `Coupon' (UCI).}
  \begin{subfigure}{0.46\textwidth}
      \scalebox{0.45}{%
%
}
    \caption{RFQ classifier.} \label{fig:coupon_rfq}
  \end{subfigure}%
  \hspace*{\fill}
  \begin{subfigure}{0.46\textwidth}
      \scalebox{0.45}{%
%
}
    \caption{BRF model.} \label{fig:coupon_brf}
  \end{subfigure}
    \caption*{\footnotesize \textbf{Note}: Adjusted minimal depth threshold (Eq. \ref{eq:min_depth_distr}). Averaged results over 100 runs.}
    \label{fig:coupon}
\end{figure}

\subsection{Validation of the method with artificial data}
\begin{figure}[ht]
\caption{Minimal depth thresholding on synthetic data.}
  \begin{subfigure}{0.46\textwidth}
    \includegraphics[width=\textwidth, trim={0.2cm 0cm 0.2cm 0.6cm},clip]{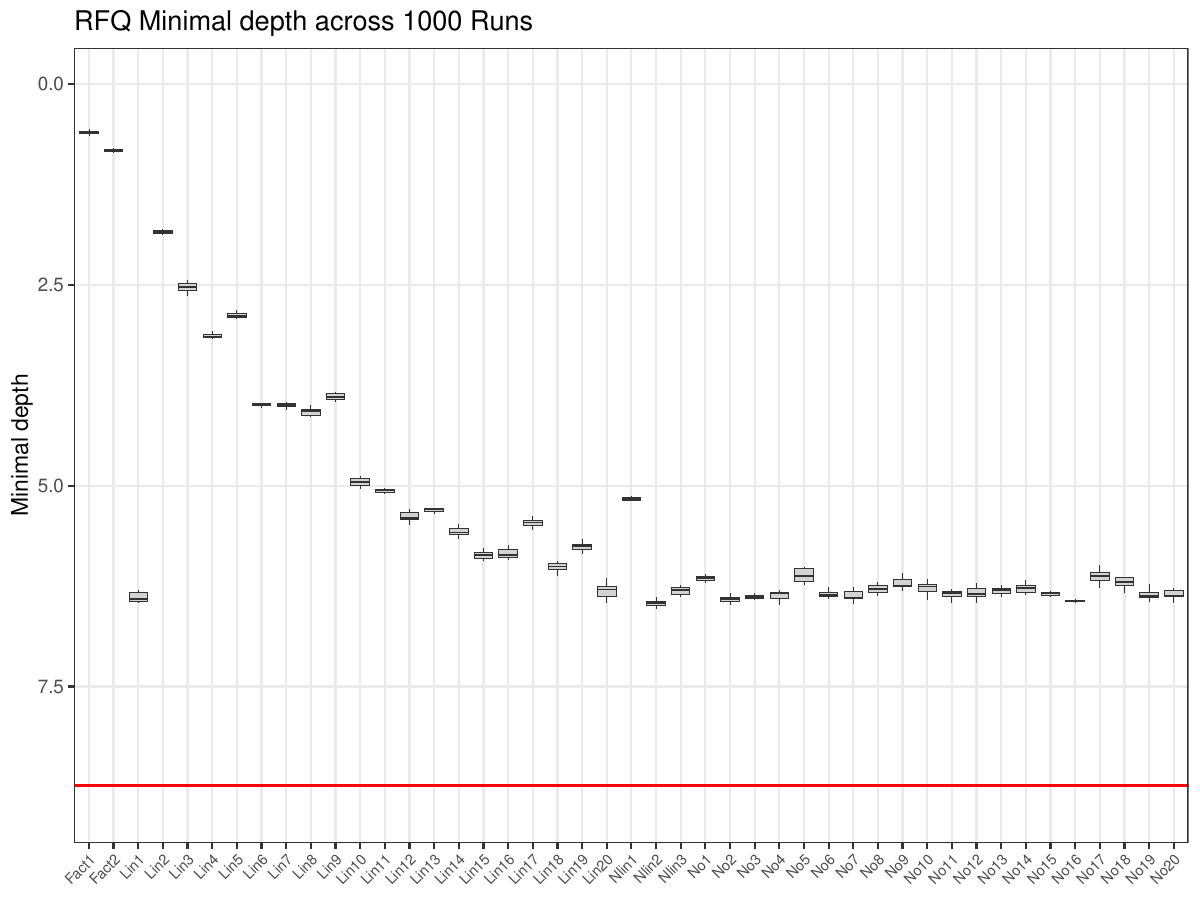}
    \caption{Minimal depth distribution and threshold defined as in \citet{Ishwaran2008}.} \label{fig:1a}
  \end{subfigure}%
  \hspace*{\fill}
  \begin{subfigure}{0.46\textwidth}
    \includegraphics[width=\textwidth, trim={0.2cm 0cm 0.2cm 0.6cm},clip]{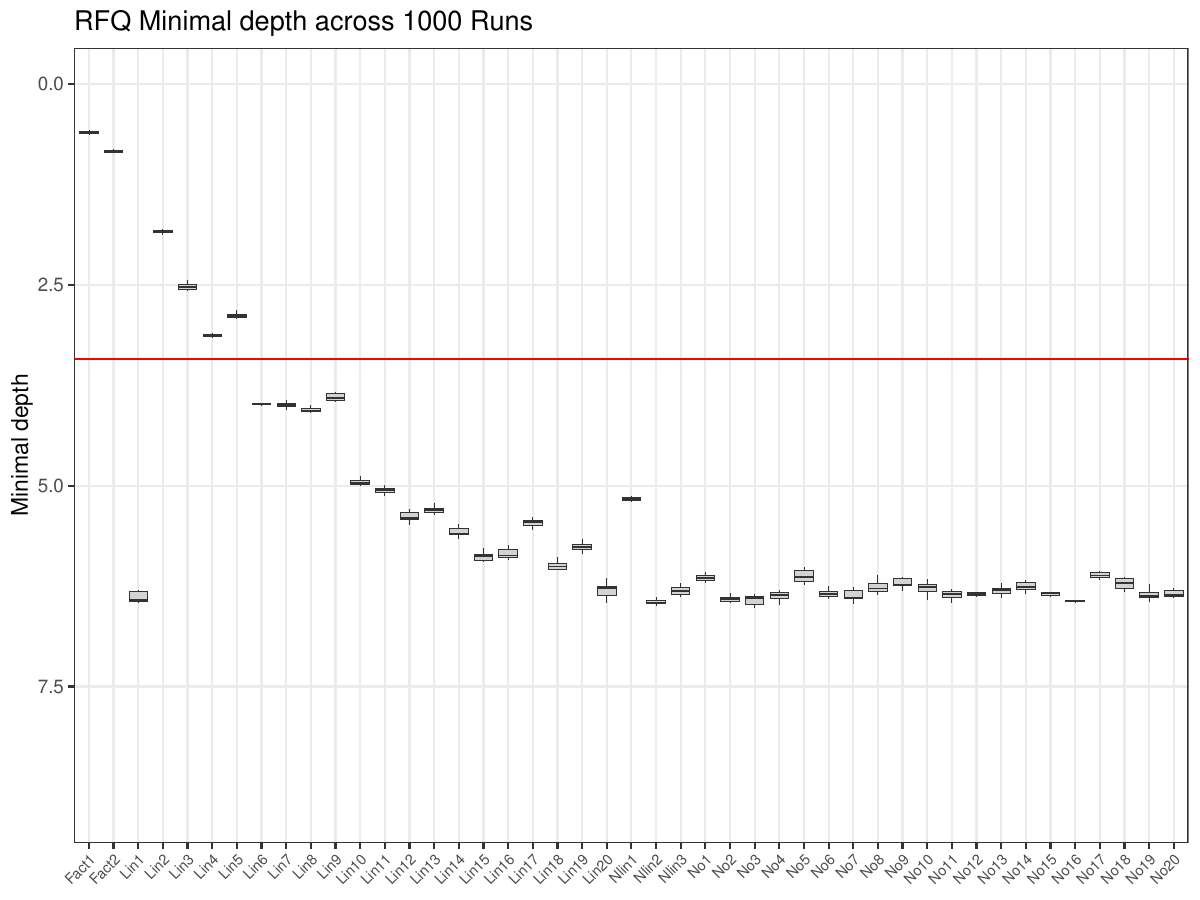}
    \caption{Proposed minimal depth distribution and threshold as in Equation \ref{eq:min_depth_distr}.} \label{fig:1b}
  \end{subfigure}
    \caption*{\footnotesize \textbf{Note}: Averaged results over 100 runs using simulated imbalanced data. There are 2 factors, 20 linear variables, 3 non-linear variables, and 20 noise variables (no signal).}
    \label{fig:mdepth_threshold}
\end{figure}

In Section \ref{sec:new_threshold}, we introduced a novel thresholding scheme for feature selection based on the mean minimal depth distribution, specifically tailored for settings where \( n \gg p \). The proposed threshold relies on the minimal depth distribution defined in Equation~\ref{eq:min_depth_distr}, incorporating a data-driven adjustment factor as specified in Equation~\ref{eq:adj_factor}.
To evaluate the performance of the proposed variable importance measures, we used the \texttt{twoClassSim} function from the \texttt{caret} package \citep{Caret} to simulate data with a binary outcome variable, two factors, 20 linear variables, 20 noise variables, and 3 non-linear variables. The sample size was set to \( n = 25{,}000 \), with an imbalance ratio (IR) of 6, achieved by downsampling \texttt{class 2}. The simulated analysis employed RFQ classifiers using forests of 5,000 trees grown on each training dataset, with \texttt{nodesize} = 1 and \texttt{mtry} = \( p/3 \). Results were averaged over 100 iterations and are presented in Figure~\ref{fig:mdepth_threshold}.

The simulation results indicate that the traditional thresholding scheme \citep{Ishwaran2008} fails to effectively discriminate relevant features, resulting in the selection of all variables in the model. In contrast, the proposed threshold successfully filters out irrelevant features, thanks to the modified minimal depth distribution.

\section{Conclusions}
\label{sec:5}
This study addressed the challenges posed by \textit{double imbalance} in supervised learning tasks, a condition characterized by both class imbalance in the target variable and a disproportionate ratio between the number of observations and features (\(n \gg p\)). Such configurations are increasingly common in large-scale applications and can lead to overfitting, biased feature selection, and reduced classification performance, especially in ensemble models like RF.
To mitigate these issues, we proposed a novel algorithm-level solution: a modified thresholding scheme for variable selection based on the minimal depth distribution. Unlike the traditional approach that relies on fixed thresholds derived from the minimal depth distribution \citep{Ishwaran2008}, our method introduces a data-driven adjustment factor that accounts for dimensional asymmetries in the dataset. This adaptation improves the reliability of feature selection by regularizing variable importance estimates in settings where deep trees can otherwise dominate.

Empirical results on both real-world and simulated datasets demonstrated the superior performance of the proposed method. In particular, our thresholding scheme consistently outperformed the classical minimal depth approach by identifying more parsimonious and informative subsets of features, without compromising classification effectiveness. The method proved especially effective in isolating relevant variables under challenging double imbalance conditions, thereby offering a robust enhancement to Random Forest-based classifiers.
Overall, this work contributes a theoretically grounded framework for feature selection in imbalanced data settings, with particular relevance for high-sample, low-dimensional contexts that are not well-served by existing techniques.

\clearpage
\bibliographystyle{elsarticle-harv} 
\bibliography{main.bib}

\end{document}